# Rotation-*k* Affine-Power-Affine-like Multiple Substitution-Boxes for Secure Communication

Musheer Ahmad
Department of Computer Engineering,
Faculty of Engineering and Technology,
Jamia Millia Islamia, New Delhi, INDIA
E-mail: musheer.cse@gmail.com

Hamed D AlSharari
Department of Electrical Engineering,
College of Engineering,
AlJouf University, AlJouf, KSA
E-mail: hamed_100@hotmail.com

*Abstract*: Substitution boxes with thorough cryptographic strengths are essential for the development of strong encryption systems. They are the only portions capable of inducing nonlinearity in symmetric encryption systems. Bijective substitution boxes having both high nonlinearities and high algebraic complexities are the most desirable to thwart linear, differential and algebraic attacks. In this paper, a method of constructing algebraically complex and cryptographically potent multiple substitution boxes is proposed. The multiple substitution boxes are synthesized by applying the concept of rotation-*k* approach on the affine-power-affine structure. It is shown that the rotation-k approach inherits all the features of affine-power-affine structure. Performance assessment of all the proposed substitution boxes is done against nonlinearity, strict avalanche criteria, bits independent criteria, differential probability, linear approximation probability and algebraic complexity. It has been found that the proposed substitution boxes have outstanding cryptographic characteristics and outperform the various recent substitution boxes.

## I. INTRODUCTION

Information security is a problem that has concerned humans since ancient times. With the explosion of the information era, information security has taken the centre-stage in today's world. Modern cryptography can be typically divided into two fields; viz. symmetric-key cryptography and asymmetric-key cryptography. Symmetric key cryptography refers to the encryption method in which the sender and the recipient share the same key. In asymmetric key cryptography, the sender and the recipient use two different keys for the purpose of encryption and decryption, respectively. In symmetric-key modern block cipher, the plaintext is broken into n-bit 'blocks', each of which is encrypted to get a cipher text block of the same length. This encryption is done using a n-bit key. Mathematically, block cipher is a mapping from the set of n-bit inputs onto the set of n-bit outputs, such that the permutation of the input-output pairs is determined by a k-bit key. By varying the key, a different permutation of the input-output pairs is obtained. The requirement of an ideal block cipher is that the relationship between the key and the permutation obtained should appear to be random [1].

Substitution boxes are indispensable nonlinear component of modern day cryptographic systems, widely used in symmetric-key encryption algorithms like DES, IDEA, AES, Blowfish, KASUMI, RC5, Lucifer, GOST, etc. The vital role of S-box is to abstruse the relationship between ciphertext and the secret key, which is also called the Shannon's property of confusion [2]. S-boxes are primarily meant to introduce nonlinearity in encryption algorithms, thereby making them resistant to linear and differential cryptanalysis [3, 4]. Thus S-box forms the core part of encryption algorithms. The strength of these algorithms is entirely dependent on the amount of confusion introduced by the S-box. The performance of the S-box is highly dependent on the area of use and also on the nature of data. Since Rinjdael S-box assumes to have excellent features and plays a critical role in the success of AES [5]. Many researchers have focused their research on evaluating, and assessing the features and strengths of AES S-box [6-9]. In [10] Cui and Cao designed a new S-box structure named affine-power-affine to improve the algebraic complexity of original Rinjdael AES S-box and making it more stronger as compared to Rinjdael S-box. But, when it comes to high auto-correlated data, as in the case of digital media like images, S-boxes show poor performance despite of having high nonlinearity. In [10-14], many attempts have been designed to construct new S-boxes similar to AES S-Box. The availability of new S-boxes is desirable in high speed communication systems while keeping the level of security same as the AES S-Box. Efficient methods are proposed to synthesize large number of S-boxes with a reasonable level of complexity. In [15], Hussain *et al.* applied the S8 permutation group to the elements of APA S-box to synthesize 40732 different S-boxes that inherits all the essential features of the APA S-box.

In this work, we proposed a simple and effective method using rotation-k approach on the elements of the famous APA S-Box. The method synthesizes multiple substitution boxes that inherit all the essential characteristics of the original APA S-Box. In Section II, the affine-power-affine structure is discussed briefly. The proposed methodology is explained in section III. The performance analyses of all the seven synthesized substitution boxes are discussed in section IV which followed by the conclusion of the work in section V.

## II. APA STRUCTURE AND CHARACTERISTICS

The AES S-box has been considered as secure against linear and differential cryptanalyses. However, AES has simple algebraic structure as $S = A \circ P$ and having only 9 terms in its algebraic expression, which makes AES S-box susceptible to algebraic attacks [6]. Cui and Cao considered the problem and introduced a new structure called Affine-Power-Affine (APA) that amplified the algebraic complexity [10]. Due to the APA structure, the algebraic complexity of improved AES S-box increases from 9 to 253. AES S-box is the combination of a power function $P(x)$ (the multiplicative inverse modulo the polynomial $x^8 + x^4 + x^3 + x + 1$) and an affine transformation $A(x)$. The affine-power-affine has the following structure

$$S(x) = A \circ P \circ A$$

Where, A represents the affine surjection and P represents inverse power permutation function over $GF(2^8)$.

$$P(x) = \begin{cases} x^{-1} & x \neq 0 \\ 0 & x = 0 \end{cases}$$

$$A(x) = \begin{bmatrix} 1 & 0 & 0 & 0 & 1 & 1 & 1 & 1 \\ 1 & 1 & 0 & 0 & 0 & 1 & 1 & 1 \\ 1 & 1 & 1 & 0 & 0 & 0 & 1 & 1 \\ 1 & 1 & 1 & 1 & 0 & 0 & 0 & 1 \\ 1 & 1 & 1 & 1 & 1 & 0 & 0 & 0 \\ 0 & 1 & 1 & 1 & 1 & 1 & 0 & 0 \\ 0 & 0 & 1 & 1 & 1 & 1 & 1 & 0 \\ 0 & 0 & 0 & 1 & 1 & 1 & 1 & 1 \end{bmatrix} \times \begin{bmatrix} x_0 \\ x_1 \\ x_2 \\ x_3 \\ x_4 \\ x_5 \\ x_6 \\ x_7 \end{bmatrix} \oplus \begin{bmatrix} 1 \\ 1 \\ 0 \\ 0 \\ 0 \\ 1 \\ 1 \\ 0 \end{bmatrix}$$

Where $x_i$'s are the coefficients of $x$ (8-bit elements of S-box in $GF(2^8)$). It has been shown by Cui and Cao that the APA structure retains the characteristics of AES S-Box as long as the power permutation P is a bijection. Let, $S: GF(2^8) \rightarrow GF(2^8)$ is multi-output mapping and $A: GF(2^8) \rightarrow GF(2^8)$ is affine surjection, then we have

$N(S) = N(A \circ S)$ and

$\delta(S) = \delta(A \circ S)$

Let, $B: GF(2^8) \rightarrow GF(2^8)$ is also an affine surjection, then

$N(S) = N(A \circ S \circ B)$ and

$\delta(S) = \delta(A \circ S \circ B)$

where, $N(S)$ and $\delta(S)$ denotes the nonlinearity and differential uniformity [10, 15]. Hence, form the above equations, we have

$N(A \circ P \circ A) = N(A \circ P)$ and

$\delta(A \circ P \circ A) = \delta(A \circ P)$

This ensures that the APA structure inherit all the cryptographic strengths and features of the AES S-Box. But, the affine operation before the power function is equivalent to multiplying with linear polynomials. As a result, APA structure improves the algebraic complexity of the S-Box, from 9 to 253, which is essential to mitigate the algebraic attacks [10].

## III. CONSTRUCTING ROTATION-*k* APA S-BOXES

The proposed multiple substitution boxes are constructed by applying the rotation-k operation on the elements of APA S-box in binary form. Firstly, all the elements of APA S-Box are converted into 8-bit binary form and then each 8-bit vector is rotated by $k$ ($0 < k < 8$) number of positions. The same rotation operation is applied on the elements to get the new 8x8 substitution box. Using the proposed transformation, seven different substitution boxes can be easily synthesized and each one has the characteristics similar to the APA substitution box. The proposed transformation can be expressed as:

$f$ : Rotation-*k*(APA-SBox) $\rightarrow$ rotation-*k* S-Box

This way, the 7 new 8x8 substitution boxes have been explored which have the algebraic complexity better than the AES S-Box. The statistical performance evaluation analyses justify this experimentally. The method of synthesizing 7 new and different substitution boxes is illustrated in Figure 1.

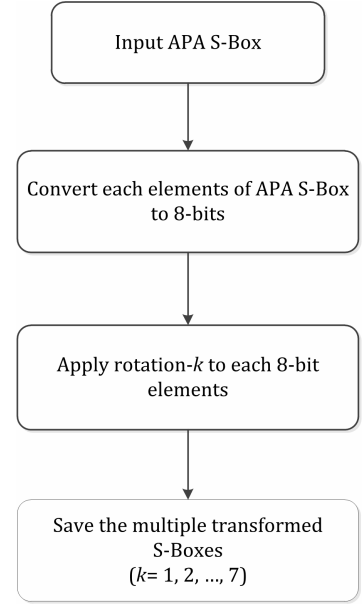

Figure 1. Construction of rotation-*k* APA S-Boxes.

## IV. STATISTICAL ANALYSIS OF ROTATION-*K* APA S-BOXES

The performance of all the rotation-k S-Boxes are analyzed and tested against various standard statistical parameters discussed in the subsequent subsections.

### A. Bijectivity

An 8x8 substitution box is said to be bijective if it has all the 256 unique elements in the [0, 255]. All the seven synthesized S-Boxes have been verified to satisfy the bijectiveness [16].

### B. Nonlinearity

Nonlinearity is an important property, which can decide the usability of an S-box as nonlinear component in block ciphers. In terms of the Walsh spectrum, it is defined as [17-25]:

$$N_g = 2^{n-1}\left(1 - 2^{-n} \max_{\omega \in GF(2^n)} |s_{\langle g \rangle}(\omega)|\right)$$

Where $N_g$ is the nonlinearity of the Boolean function and the Walsh spectrum of $g(x)$ is described as:

$$S_{\langle g \rangle}(\omega) = \sum_{x \in GF(2^n)} (-1)^{g(x) \oplus x \bullet \omega}$$

Where $\omega$ belongs to $GF(2^8)$ and $x.\omega$ denotes the scalar product of $x$ and $\omega$. High nonlinearity scores of all eight Boolean functions in S-boxes are requisite since it diminishes the input-output correlation. Following the mathematics for proposed 8x8 S-box, the nonlinearity of eight Boolean functions ($1 \leq g_i \leq 8$) involved are evaluated. The nonlinearity performances of the seven S-boxes are tested and compared with some recent 8x8 S-boxes in Table II. Comparatively, the synthesized S-boxes have excellent nonlinearity strength and outperform many recent S-boxes.

### C. Strict Avalanche Criteria

The strict avalanche criterion (SAC) was initially presented by Webster and Tavares [4] in 1986. Mathematically, a Boolean function satisfies SAC criteria if each of its output bits change with a probability of a half whenever a single input bit *x* is complemented. Generally, the dependency matrix used to test the SAC of an S-box. If each element and the mean value of the matrix are both close to the ideal value of 0.5, the S-box is said to have nearly fulfilled the SAC criterion [19]. The value of SAC for the generated S-box is 0.5007 which is very close to the ideal value 0.5 SAC of proposed S-boxes and others are provided in Table IV. Moreover, the comparisons drawn in Table highlight that the proposed S-box has relevant and comparable value with respect to strict avalanche criteria.

### D. Differential Uniformity

A poor S-box design is easily vulnerable to the differential cryptanalysis. To avoid such scenarios, S-boxes should ideally have maximum value of differential uniformity as low as possible. To ensure a uniform mapping probability, an input differential should map uniquely to an output differential for each j. The differential approximation probability, for an S-box, is a measure of differential uniformity [26] which is defined as:

$$DP(\delta x \rightarrow \delta y) = (\#\{x \in X | f(x) \oplus f(x \oplus \delta x) = \delta y\})$$

Here, $X$ is the set of all input values and $2^n$ are number of S-box elements. The maximum value of DU should be as low as possible for a strong substitution-box to thwart the differential attacks. The maximum differential probability, listed in Table IV, for the proposed S-box is 4, which is best than the maximum DP of S-boxes investigated by Jakimoski, Khan, Khan and Gondal *et al.* in [27, 29-31].

Table I. Proposed rotation-*k* (for *k* = 3) APA substitution box

|    | 1   | 2   | 3   | 4   | 5   | 6   | 7   | 8   | 9   | 10  | 11  | 12  | 13  | 14  | 15  | 16  |
|----|-----|-----|-----|-----|-----|-----|-----|-----|-----|-----|-----|-----|-----|-----|-----|-----|
| 1  | 145 | 27  | 105 | 184 | 161 | 138 | 227 | 107 | 155 | 185 | 235 | 129 | 76  | 158 | 156 | 215 |
| 2  | 18  | 20  | 38  | 164 | 237 | 43  | 118 | 21  | 193 | 202 | 68  | 154 | 110 | 192 | 19  | 128 |
| 3  | 59  | 51  | 182 | 135 | 172 | 251 | 125 | 104 | 122 | 73  | 97  | 149 | 121 | 213 | 127 | 7   |
| 4  | 56  | 211 | 90  | 49  | 245 | 229 | 249 | 58  | 165 | 218 | 10  | 64  | 40  | 186 | 89  | 163 |
| 5  | 200 | 24  | 98  | 57  | 82  | 91  | 29  | 47  | 77  | 1   | 167 | 2   | 11  | 220 | 34  | 195 |
| 6  | 108 | 178 | 39  | 101 | 244 | 148 | 232 | 176 | 75  | 112 | 16  | 144 | 46  | 103 | 190 | 94  |
| 7  | 106 | 236 | 141 | 71  | 222 | 160 | 74  | 72  | 136 | 119 | 67  | 207 | 238 | 87  | 187 | 228 |
| 8  | 62  | 246 | 180 | 88  | 212 | 146 | 198 | 80  | 117 | 102 | 247 | 45  | 131 | 223 | 79  | 35  |
| 9  | 44  | 173 | 96  | 205 | 8   | 115 | 151 | 248 | 25  | 60  | 153 | 14  | 111 | 210 | 139 | 86  |
| 10 | 70  | 28  | 231 | 216 | 55  | 78  | 194 | 52  | 188 | 6   | 255 | 12  | 241 | 252 | 191 | 174 |
| 11 | 162 | 126 | 169 | 85  | 189 | 32  | 37  | 15  | 230 | 201 | 140 | 170 | 243 | 225 | 217 | 84  |
| 12 | 199 | 5   | 134 | 50  | 22  | 142 | 206 | 233 | 65  | 132 | 240 | 69  | 179 | 168 | 17  | 221 |
| 13 | 196 | 4   | 63  | 41  | 120 | 53  | 66  | 92  | 109 | 203 | 99  | 36  | 116 | 133 | 26  | 123 |
| 14 | 83  | 208 | 157 | 3   | 250 | 254 | 95  | 166 | 42  | 150 | 152 | 234 | 54  | 30  | 13  | 23  |
| 15 | 242 | 214 | 209 | 114 | 175 | 48  | 147 | 93  | 124 | 0   | 224 | 100 | 239 | 159 | 177 | 33  |
| 16 | 197 | 253 | 226 | 219 | 143 | 61  | 81  | 181 | 130 | 9   | 31  | 204 | 171 | 183 | 137 | 113 |

Table II. Nonlinearities of some 8x8 substitution boxes

| | Substitution-Boxes | 1 | 2 | 3 | 4 | 5 | 6 | 7 | 8 | Average |
|---|---|---|---|---|---|---|---|---|---|---|
| **P R O P O S E D** | *rotation*-1 APA S-Box | 112 | 112 | 112 | 112 | 112 | 112 | 112 | 112 | 112 |
| | *rotation*-2 APA S-Box | 112 | 112 | 112 | 112 | 112 | 112 | 112 | 112 | 112 |
| | *rotation*-3 APA S-Box | 112 | 112 | 112 | 112 | 112 | 112 | 112 | 112 | 112 |
| | *rotation*-4 APA S-Box | 112 | 112 | 112 | 112 | 112 | 112 | 112 | 112 | 112 |
| | *rotation*-5 APA S-Box | 112 | 112 | 112 | 112 | 112 | 112 | 112 | 112 | 112 |
| | *rotation*-6 APA S-Box | 112 | 112 | 112 | 112 | 112 | 112 | 112 | 112 | 112 |
| | *rotation*-7 APA S-Box | 112 | 112 | 112 | 112 | 112 | 112 | 112 | 112 | 112 |
| **E X I S T I N G** | APA S-Box [10] | 112 | 112 | 112 | 112 | 112 | 112 | 112 | 112 | 112 |
| | $S_8$ APA S-Box [15] | 112 | 112 | 112 | 112 | 112 | 112 | 112 | 112 | 112 |
| | AES S-Box [5] | 112 | 112 | 112 | 112 | 112 | 112 | 112 | 112 | 112 |
| | Skipjack S-Box | 104 | 104 | 108 | 108 | 108 | 104 | 104 | 106 | 105.75 |
| | In [27] S-Box | 98 | 100 | 100 | 104 | 104 | 106 | 106 | 108 | 103.25 |
| | In [28] S-Box | 104 | 100 | 106 | 102 | 104 | 102 | 104 | 104 | 103.25 |
| | In [29] S-Box | 108 | 102 | 100 | 104 | 104 | 102 | 98 | 106 | 103 |
| | In [30] S-Box | 100 | 108 | 106 | 104 | 102 | 102 | 106 | 108 | 104.5 |
| | In [31] S-Box | 98 | 100 | 106 | 104 | 106 | 100 | 106 | 104 | 103 |

## E. Bit Independent Criteria

The bit independence criterion is given by Webster and Tavares which deals with testing an individual bit at the input of the cipher by performing the toggle operation. The bit independence criterion (BIC) analyzes all the avalanche variables and determines the extent of their pair-wise independence in reference to a given set of avalanche vectors. The avalanche vectors are generated by the bit patters resulting from complementing bit(s) at the input according to Webster & Tavares [4]. It is a desirable property for any cryptographic design. It means that all the avalanche variables should be pair-wise independent for a given set of avalanche vectors generated by complementing a single plaintext bit. In order to measure the degree of independence between a pair of avalanche variables, we can calculate their correlation coefficient. For two variables A and B,

$$\rho(A,B) = \frac{\text{cov}(A,B)}{\sigma(A)\sigma(B)}$$

Where $\rho(A, B)$ is the correlation coefficient of A and B, $cov(A, B)$ is the covariance of A and B, i.e. cov(A, B) = E(AB) – E(A) × E(B) and $\sigma^2(A) = E(A^2) - (E(A))/2$. Suppose, the Boolean functions in the 8×8 S-box are $f_1, f_2, \ldots f_8$. It was pointed out that if the S-box met BIC, $f_j \oplus f_k$ ($j \neq k$, $1 \leq j, k \leq 8$) should be highly nonlinear and satisfies the avalanche criterion [17, 28]. Therefore, BIC can be verified by calculating the SAC and the nonlinearity of $f_j \oplus f_k$. The BIC nonlinearity scores of $f_j \oplus f_k$ are depicted in Table III for the proposed S-box. The average of BIC scores for nonlinearity are quantified and listed in Table III. The average of bits independent criteria with respect to nonlinearities is 112. The BIC scores justify the satisfactory performance of proposed S-box.

Table III. Bit independent criteria for nonlinearity for proposed substitution-box

| -   | $f_1$ | $f_2$ | $f_3$ | $f_4$ | $f_5$ | $f_6$ | $f_7$ | $f_8$ |
|-----|-----|-----|-----|-----|-----|-----|-----|-----|
| $f_1$ | 0   | 112 | 112 | 112 | 112 | 112 | 112 | 112 |
| $f_2$ | 112 | 0   | 112 | 112 | 112 | 112 | 112 | 112 |
| $f_3$ | 112 | 112 | 0   | 112 | 112 | 112 | 112 | 112 |
| $f_4$ | 112 | 112 | 112 | 0   | 112 | 112 | 112 | 112 |
| $f_5$ | 112 | 112 | 112 | 112 | 0   | 112 | 112 | 112 |
| $f_6$ | 112 | 112 | 112 | 112 | 112 | 0   | 112 | 112 |
| $f_7$ | 112 | 112 | 112 | 112 | 112 | 112 | 0   | 112 |
| $f_8$ | 112 | 112 | 112 | 112 | 112 | 112 | 112 | 0   |

## F. Linear Approximation Probability

The linear approximation probability (LAP) is the maximum value of the imbalance of an event. The parity of the input bits selected by the mask $\Gamma x$ is equal to the parity of the output bits selected by the mask $\Gamma y$. A linear approximation probability of the likelihood (or probability bias) of the S-box is defined as [32]:

$$LAP = \max_{\Gamma x, \Gamma y \neq 0y} \left( \frac{\#\{x \in X \mid x \bullet \Gamma x = f(x) \bullet \Gamma x\}}{2^n} \right)$$

Where, $\Gamma x$ and $\Gamma y$ are input and output masks, respectively; $X$ is the set of all possible inputs; and $2^n$ is the number of its elements. Like DU, It should also be as small as possible for a strong substitution-box. The linear approximation probability of proposed S-boxes is found as 0.0625. Hence, it can be said that the proposed S-box when used for block encryption and can offer resistance to the linear approximation attacks, thereby, providing the security strength to the cryptosystem.

## G. Algebraic Complexity

S-boxes are traditionally based of power mappings of the form $x^d$ for some exponent $d$. In the case of the AES, Fermat's Little Theorem tells us that $d = 254 = -1$ over $GF(2^8)$ [12]. The inverse power mapping inside this S-box is then augmented with the affine transformation. The algebraic complexity is defined as the number of terms in the linearized polynomial. So, for the AES, the algebraic complexity is equal to 9. Some researchers fear that this is too low and may render variations of interpolation attacks successful [14]. As such, there has been ample work done to increase the algebraic complexity to higher values. Here we will discuss how to apply Lagrangian interpolation to find algebraic complexity of an S-box [33]. As Daemen and Rijmen pointed out, any function from a finite field to itself can be expressed as a polynomial. In fact, given a tabular form of the function, it is possible to generate the Lagrange polynomial and then simplify. So let's start with Lagrange polynomial [34].

It is experimentally verified that the algebraic complexity for all 7 synthesized S-boxes is comes out as 253 which is quite excellent than complexity 9 of AES S-box and similar to the features of the APA S-box and $S_8$ APA S-boxes as listed in Table V.

Table IV. SAC, max DU, average BIC-NN and max LP of some substitution-boxes

| S-Box | SAC | Max DU | Average BIC-NN | Max LP |
|---|---|---|---|---|
| *rotation*-1 APA S-Box | 0.5007 | 4 | 112 | 0.0625 |
| *rotation*-2 APA S-Box | 0.5007 | 4 | 112 | 0.0625 |
| *rotation*-3 APA S-Box | 0.5007 | 4 | 112 | 0.0625 |
| *rotation*-4 APA S-Box | 0.5007 | 4 | 112 | 0.0625 |
| *rotation*-5 APA S-Box | 0.5007 | 4 | 112 | 0.0625 |
| *rotation*-6 APA S-Box | 0.5007 | 4 | 112 | 0.0625 |
| *rotation*-7 APA S-Box | 0.5007 | 4 | 112 | 0.0625 |
| APA S-Box [10] | 0.5007 | 4 | 112 | 0.0625 |
| $S_8$ APA S-Box [15] | 0.5007 | 4 | 112 | 0.0625 |
| AES S-Box [5] | 0.504 | 4 | 112 | 0.0625 |
| Skipjack S-Box | 0.503 | 12 | 104.14 | 0.109 |
| In [27] S-Box | 0.4972 | 12 | 104.2 | 0.1289 |
| In [28] S-Box | 0.5048 | 10 | 103.7 | 0.1289 |
| In [29] S-Box | 0.5012 | 12 | 104.1 | 0.1484 |
| In [30] S-Box | 0.4978 | 12 | 103.6 | 0.1406 |
| In [31] S-Box | NR | 12 | 104.14 | 0.1484 |

Table V. Algebraic complexities

| S-Box | Algebraic Complexity |
|---|---|
| *rotation-k* APA S-Box (all 7) | 253 |
| APA S-Box [10] | 253 |
| $S_8$ APA S-Box [5] | 253 |
| AES S-Box [5] | 9 |

## V. CONCLUSION

In this paper, we proposed a method of constructing algebraically complex and cryptographically potent multiple substitution boxes that have strength as that of the original affine-power-affine S-Box. The seven substitution boxes are constructed using the concept of rotation-k operation on the elements of APA S-Box. It has been shown that all the rotation-k approach-based seven S-Boxes inherit all the cryptographic features of affine-power-affine structure. Performance assessment of all new seven substitution boxes is carried out against worldwide accepted standard measures like nonlinearity, strict avalanche criteria, bits independent criteria, differential probability, linear approximation probability and algebraic complexity. Moreover, it has also been shown that the proposed substitution boxes have excellent cryptographic strengths and outperform various recent substitution boxes. The new seven substitution boxes are perfectly suitable candidates for the design of block encryption systems to realize secure communications.

## VI. REFERENCES

[1] B. Schneier, Applied Cryptography: Protocols Algorithms and Source Code in C. New York, Wiley, 1996.

[2] C.E. Shannon, "Communication theory of secrecy systems" Bell system technical journal, vol. 28 (4), pp. 656-715, 1949.

[3] L. Keliher, "Refined analysis of bounds related to linear and differential and linear cryptanalysis for the AES", In: Dobbertin H, et al., editors. Advanced Encryption Standard–AES, Lecture notes in computer science, pp. 42–57, 2005.

[4] A. Webster, S. Tavares, "On the design of S-boxes", In: Advances in cryptology: Proceedings of CRYPTO'85. Lecture notes in computer science, pp. 523–534, 1986.

[5] J. Daemen and V. Rijmen, AES Proposal: Rijndael, http://www.east.kuleuven.ac.be/ rijmen/ rijndael, 1999.

[6] M.T. Sakallı, B. Aslan, E. Buluş, A.S., Mesut, F. Büyüksaraçoğlu, O. Karaahmetoğlu, "On the Algebraic Expression of the AES S-Box Like S-Boxes", In: Networked Digital Technologies, Springer Berlin Heidelberg , pp. 213-227, 2010.

[7] N. Ferguson and J. Kelsey, "Improved cryptanalysis of Rijndael", The 7th International Workshop of Fast Software Encryption, New York, pp.213-230, 2001.

[8] S. Murphy and M. J. B. Robshaw, "Essential algebraic structure within the AES", Proceedings of the 22nd Annual International Cryptology Conference on Advances in Cryptology, pp.1-16, 2002.

[9] J. Detombe, S. Tavares, "Constructing large cryptographically strong S-boxes", Advances in cryptology: Proceedings of AUSCRYPT'1992, Lecture notes in computer science, vol. 718, pp. 165-181, 1992.

[10] L. Cui and Y. Cao, "A new S-box structure named Affine-Power-Affine", International Journal of Innovative Computing, Information and Control, vol. 3(3), pp. 2007.

[11] M.T. Tran, D.K. Bui, and A.D. Duong, "Gray sbox for advanced encryption standard", International Conference on Computational Intelligence and Security, pp. 2008.

[12] J. Liu, B. Wai, X. Cheng, and X. Wang. An AES S-box to increase complexity and cryptographic analysis. In Proceedings of the 19th International Conference on Advanced Information Networking and Applications, vol. 1, pp. 724–728, 2005.

[13] B.N. Tran, T.D. Nguyen, T.D. Tran, "A new S-box structure to increase complexity of algebraic expression for block cipher cryptosystems", In: International Conference on Computer Technology and Development, Vol. 2, pp. 212-216, 2009.

[14] J. Cui, L. Huang, H. Zhong, C. Chang, W. Yang, "An improved AES S-Box and its performance analysis" International Journal of Innovative Computing, Information and Control, vol. 7(5), pp. 2291-2302, 2011.

[15] I. Hussain, T. Shah, M.A. Gondal, H. Mahmood, "$S_8$ affine-power-affine S-boxes and their applications", Neural Computing and Application, vol. 21, pp. S377–S383, 2012.

[16] M. Ahmad, D. Bhatia, Y. Hassan, "A Novel Ant Colony Optimization Based Scheme for Substitution Box Design." Procedia Computer Science 57, pp. 572-580, 2015.

[17] D. Lambić, "A novel method of S-box design based on chaotic map and composition method", Chaos, Solitons & Fractals, vol. 58, pp. 16-21, 2014.

[18] T. Cusick, P. Stanica, "Cryptographic boolean functions and applications", Elsevier, 2009.

[19] M. Ahmad, H. Chugh, A. Goel, P. Singla, "A chaos based method for efficient cryptographic S-box design", International Symposium on Security in Computing and Communications, CCIS, vol. 377, pp. 130-137, 2013.

[20] M. Ahmad, S. Alam, "A Novel Approach for Efficient S-Box Design Using Multiple High-Dimensional Chaos." International Conference on Advanced Computing & Communication Technologies, pp. 95-99, 2014.

[21] M. Ahmad, P. M. Khan, M. Z. Ansari, "A simple and efficient keydependent S-box design using Fisher-Yates shuffle technique", International Conference on Security in Networks and Distributed Systems, CCIS, vol. 420, pp. 540-550, 2014.

[22] M. Ahmad, H. Haleem, P.M. Khan, "A new chaotic substitution box design for block ciphers." International Conference on Signal Processing and Integrated Networks, pp. 255-258, 2014.

[23] M. Ahmad, F. Ahmad, Z. Nasim, Z. Bano, S. Zafar, "Designing chaos based strong substitution box." International Conference on Contemporary Computing, pp. 97-100, 2015.

[24] M. Ahmad, D.R. Rizvi, Z. Ahmad, "PWLCM-Based Random Search for Strong Substitution-Box Design", Second International Conference on Computer and Communication Technologies, pp. 471-478. Springer, 2016.

[25] P. K. Sharma, M. Ahmad, P.M. Khan, "Cryptanalysis of image encryption algorithm based on pixel shuffling and chaotic S-box transformation", In: Security in Computing and Communications, Springer Berlin Heidelberg, pp. 173-181, 2014.

[26] E. Biham and A. Shamir, "Differential cryptanalysis of DES-like cryptosystems", Journal of Cryptology, vol 4, no. 1, pp. 3-72, 1991.

[27] G. Jakimoski, and L. Kocarev, "Chaos and cryptography: Block encryption ciphers based on chaotic maps", IEEE Transaction on Circuits Systems, vol. 48, no. 2, pp. 163-169, 2001.

[28] F. Özkaynak and A. B. Özer, "A method for designing strong S-boxes based on chaotic Lorenz system", Physics Letters A, vol. 374, no. 36, pp. 3733–3738, 2010.

[29] M. Khan, T. Shah, H. Mahmood and M. A. Gondal, "An efficient method for the construction of block cipher with multi-chaotic systems", Nonlinear Dynamics, vol. 71, no. 3, pp. 489-492, 2013

[30] M. Khan and T. Shah, "An efficient construction of substitution box with fractional chaotic system", Signal, Image and Video Processing, vol. , no. , pp. , November 2013. doi 10.1007/s11760-013-0577-4.

[31] M. A. Gondal, A. Raheem, I. Hussain, "A Scheme for Obtaining Secure S-Boxes Based on Chaotic Baker's Map", 3D Research, pp. 5-17, 2014. doi: 10.1007/s13319-014-0017-4

[32] M. Matsui, "Linear Cryptanalysis Method of DES Cipher", Advances in Cryptology: EuroCrypt'1993 Proceedings, Lecture Notes in Computer Science, vol. 765, pp. 386-397, 1994.

[33] C. Cid, S. Murphy, M. Robshaw, "Computational and algebraic aspects of the advanced encryption standard", In: Proceedings of the Seventh International Workshop on Computer Algebra in Scientific Computing, 2004.

[34] J. Liu, W. Bao-dian, C. Xiang-guo, W. Xin-mei, "Cryptanalysis of Rijndael S-box and improvement", Applied Mathematics and Computation, vol. 170, no. 2, pp. 958-975, 2005.